\RequirePackage{lineno}
\documentclass[aps,prc,superscriptaddress, showpacs]{revtex4}

\usepackage{epsfig}
\usepackage{amssymb}
\usepackage{bm}
\usepackage{graphicx,color}
\usepackage{subfig}
\usepackage{dcolumn}
\usepackage{epstopdf}
\usepackage{csquotes}
\usepackage{amsmath}
\usepackage{xfrac}

\begin{document}

%\begin{frontmatter}
%\setpagewiselinenumbers
%\modulolinenumbers[5]

%\linenumbers

\title{Imaging of PbWO$_{4}$ Crystals for G Experiment Test Masses Using a Laser Interferometer}
\author {K. T. A. Assumin-Gyimah}
\affiliation{Mississippi State University, Mississippi State, MS 39762, USA}
\author {M. G. Holt}
\affiliation{Mississippi State University, Mississippi State, MS 39762, USA}
\author{D. Dutta}
\affiliation{Mississippi State University, Mississippi State, MS 39762, USA}
\author {W. M. Snow}
\affiliation{Indiana University/CEEM, 2401 Milo B. Sampson Lane, Bloomington, IN 47408, USA}

%\date{}

\begin{abstract}

It is highly desirable for future measurements of Newton's gravitational constant $G$ to use test/source masses that allow nondestructive, quantitative internal density gradient measurements. High density optically transparent materials are ideally suited for this purpose since their density gradient can be measured with laser interferometry, and they allow in-situ optical metrology methods for the critical distance measurements often needed in a $G$ apparatus. We present an upper bound on possible internal density gradients in lead tungstate (PbWO$_4$) crystals determined using a laser interferometer. We placed an upper bound on the fractional atomic density gradient in two PbWO$_{4}$ test crystals of ${1 \over \rho}{d\rho \over dx}<2.1 \times 10^{-8}$ cm$^{-1}$. This value is more than two orders of magnitude smaller than what is required for $G$ measurements. They are also consistent with but more sensitive than a recently reported measurements of the same samples, using neutron interferometry. These results indicate that PbWO$_4$ crystals are well suited to be used as test masses in $G$ experiments. Future measurements of internal density gradients of test masses used for measurements of $G$ can now be conducted non-destructively for a wide range of possible test masses. 

\end{abstract}

\maketitle

\section{Introduction}

\begin{figure}[h!]
  \begin{center}
  \includegraphics[width=16cm]{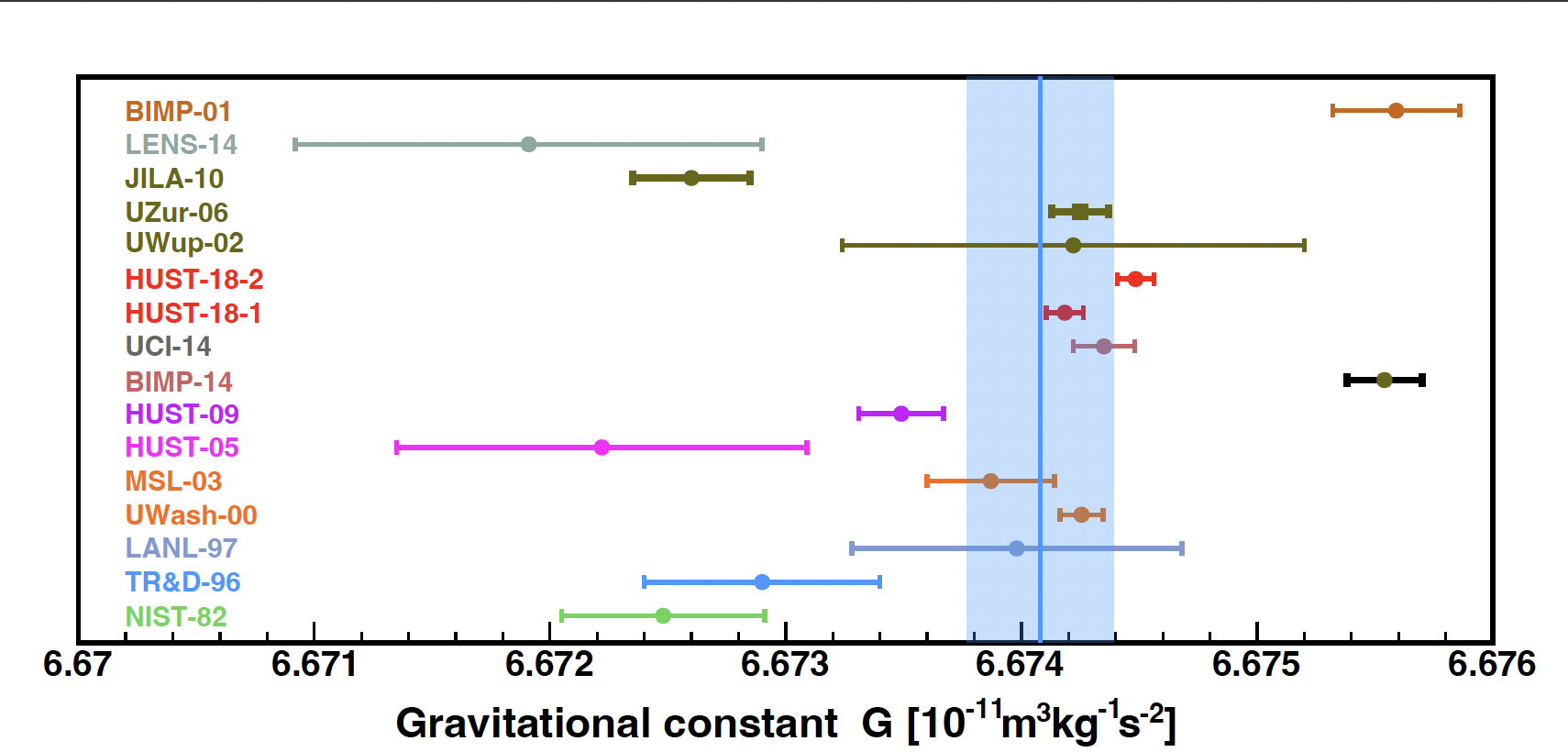}
  \caption{\textit{The most precise measurements of $G$ over the last 30 years. The blue line and band shows the 2018 CODATA recommended value. The red data points come from two new recent measurements of G~\cite{Li2018}, since the 2018 CODATA evaluation.}}
  \label{fig:fig1}
  \end{center}
\end{figure}

Gravity is universal but also the weakest  of the four fundamental forces and it is impossible to shield. The gravitational force between laboratory size objects tend to be too small to measure accurately. Consequently, the precision determination of the gravitational constant $G$, is a serious challenge. The first measurement of the universal gravitational constant $G$ was performed by Henry Cavendish~\cite{cavendish} in 1798. Despite a long history of ever improving subsequent measurements, our current knowledge
of $G$  is unusually poor relative to other fundamental constants.  The relative uncertainty of the most recent (2018) recommended value by the Committee on Data for Science and Technology (CODATA)~\cite{codata19} is $22 \times 10^{-6}$. The most precise measurements conducted in the past 30 years that were used in the 2018 CODATA evaluation are shown in Figure~\ref{fig:fig1} along with the two most recent measurements~\cite{Li2018}.  The scatter among the values from different experiments (500 parts per million) are significantly larger than what is expected from the quoted errors ($\sim$10 parts per million). This scatter among the measured values may be due to lack of complete understanding of either the systematics used to perform these measurements, or the physics behind gravitation, or both. 
Typical $G$ experiments measure small forces,
torques, or accelerations with a relative uncertainty of about $10^{-5}$, and it is likely that the scatter in this data represents unaccounted-for systematic errors which plague the metrology of small forces. These metrological issues have been discussed at length in a recent review~\cite{Rothleitner2017}. 
It is also likely that general relativity, the currently accepted description of gravitation, is not complete since a successful unification of gravitation with quantum mechanics remains elusive~\cite{grreview}. 
%The scatter in the measured values of $G$ shown in Fig.~\ref{fig:fig1} could be an indication of a much richer field of gravity. 

Given this situation, future precision $G$ measurements will justifiably be held to a
higher standard for their analysis and quantitative characterization of systematic errors.
The procedures for corrections to the raw $G$ data from apparatus
calibration and systematic errors using subsidiary measurements is very specific to the particular measurement apparatus and approach. However, there are some sources of systematic uncertainties that are common to almost all precision measurements of $G$, for example the metrology of the source/test masses used in the experiments~\cite{bigGreview}. A research program which successfully addresses this issue can help improve G measurements. 

Past measurements of $G$ have employed a variety of strategies to measure the critical dimensions and constrain the
effects of internal mass density gradients.  The source and test masses used in previous experiments have been manufactured using optically-opaque metals and alloys which might possesses hidden density inhomogeneities. 
%The most common technique used to probe such inhomogeneities is to measure small sections and/or pieces cut from same stock of material used to manufacture the source and test masses using ultrasound or xray and gamma ray radiography and tomography. Small-scale density nonuniformities from internal voids or inclusions have been constrained with such measurements. 
Systematics due to density nonuniformities such as a linear density gradient have been constrained in the past using destructive evaluation of a subset of a series of nominally identical masses. Such destructive evaluations have shown fractional density gradients large enough to prevent 1ppm $G$ measurements.  In some cases, any large-scale linear density gradients are indirectly constrained by mounting the mass on a low-friction platform and measuring the frequency of oscillation of the test/field mass considered as a physical pendulum, which determines the relevant offset between the center of mass and the symmetry axes of the mass.

We aim to characterize optically transparent masses for $G$ experiments. The use of transparent source/test masses in $G$ experiments enables nondestructive, quantitative internal density gradient measurements using laser interferometry and can help prepare the way for optical metrology methods for the critical distance measurements needed in many $G$ measurements. The density variations of glass and single crystals are generally much smaller than those for metals~\cite{gillies2014} and hence they are likely to be a better choice for source/test masses for experiments which will need improved systematic errors. It is also desirable to use a transparent test mass with a large density.

A typical $G$ measurement instrument involves distances on the order of 50~cm and source mass dimensions on the order of
10~cm. The masses are usually arranged in a pattern which lowers the sensitivity of the gravitational signal to
small shifts $\delta R$ in the location of the true center of mass from the geometrical center of the masses by about two orders of magnitude. Under these typical conditions, we conclude that in order to attain a 1 ppm uncertainty in mass metrology, the internal number density gradients of the test masses must be controlled at the level of about ${1 \over N} {dN \over dx}=10^{-4}$/cm and the absolute precision of the distance measurement to sub-micron precision. It is also valuable to use a test mass with a high density to maximize the gravitational signal strength while also bringing the masses as close as possible to each other consistent with the other experimental constraints. 

We have chosen to characterize density gradients in lead tungstate (PbWO$_{4}$) to the precision required for ppm-precision $G$ measurements. We have performed this characterization using both laser interferometry and neutron interferometry. This paper discusses the results of the characterization using laser interferometry.

\section{Relevant Properties of PbWO$_{4}$}

\begin{figure}[h!]
  \begin{center}
  \includegraphics[width=9cm]{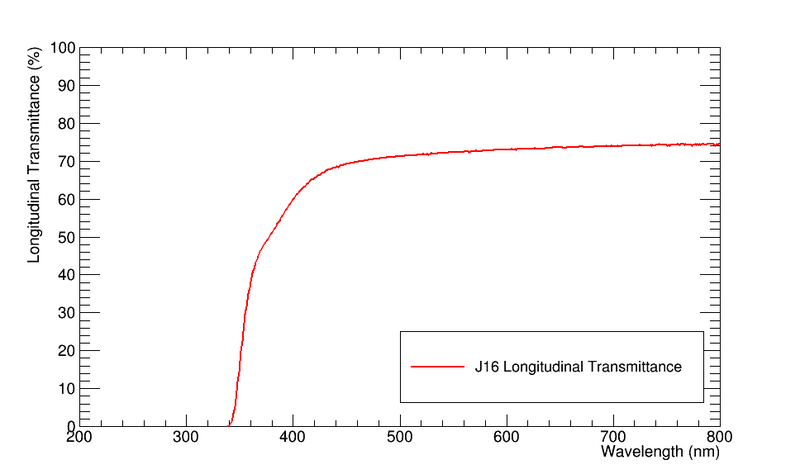}
  \includegraphics[width=8cm]{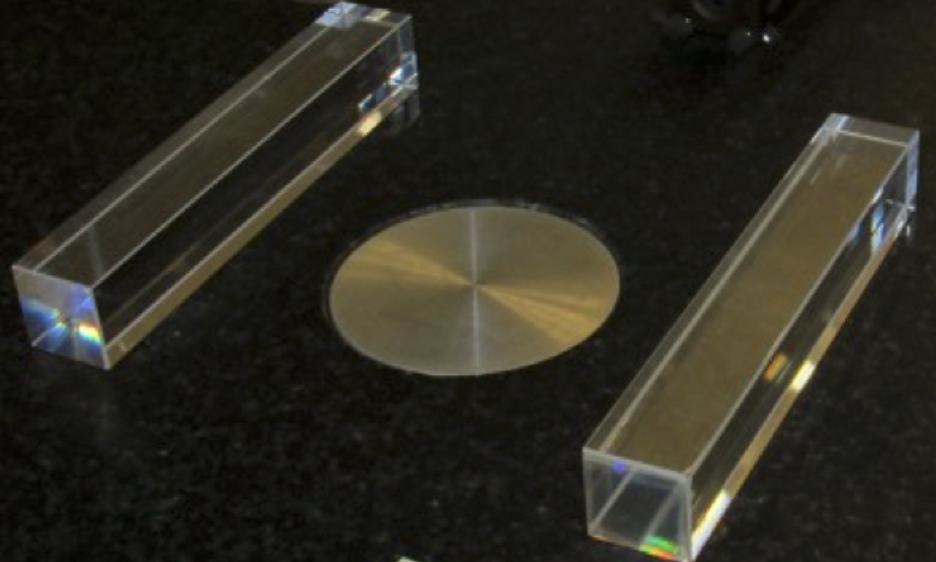}
  \caption{(left) Transmittance of PbWO$_{4}$ as a function of wavelength of incident light. (right) The two  2.3 cm x 2.3 cm x 12.0 cm PbWO$_{4}$ samples used in the optical measurement.}
  \label{fig:transp}
  \end{center}
\end{figure}

Lead tungstate is dense, non-hygroscopic, optically transparent, nonmagnetic and their internal density gradients can be characterize by optical techniques.
As shown in Fig.~\ref{fig:transp}~(left), lead tungstate is transparent for the entire visible spectrum. They have been developed for high energy physics as a high Z scintillating crystal. The low impurity concentrations and high density uniformity that were developed to meet the technical requirements for efficient transmission of the internal scintillation light inside these crystals also match the requirements for $G$ test and field masses. Lead tungstate can be grown in very large optically transparent single crystals in the range of sizes needed for $G$ experiments and are machinable to the precision required to determine $G$. These crystals are commercially available, for example from the Shanghai Institute of Ceramics (SIC) and currently several tons of PbWO$_{4}$ are being use in nuclear and high energy physics experiments all over the world. It is also being actively studied for several new detectors under construction and for R\&D on future detectors to be used at the recently-approved Electron Ion Collider. We therefore foresee a long-term motivation for continued R\&D on crystal quality and size from nuclear and high energy physics. The SIC grows boules of PbWO$_{4}$ measuring 34 mm$\times$34 mm$\times$360 mm, which are then diamond cut and polished to the desired size (typically 24 mm$\times$24 mm$\times$260 mm)~\cite{sic1}. SIC has recently produced crystals of 60~mm in diameter and are capable of producing crystals with a diameter of 100~mm and length of 320~mm~\cite{sic2}. These larger sizes are very suitable for $G$ experiments~\cite{decca}. 

Great effort is put into minimizing or eliminating common impurities such as Mo$^{6+}$, Fe$^{2+}$, Na$^{+}$, K$^{+}$ and Y$^{3}$ in PbWO$_{4}$ as they affect the crystal quality, degrade the optical transmittance properties, reduce scintillation light yield, and produce radiation damage.  Analysis using glow discharge mass spectroscopy (GDMS) indicates that most of the common impurities can be reduced to $<$ 1 ppm by weight~\cite{sic1}. The largest impurity, at 32 ppm by weight, is Y$^{3+}$ which has a direct effect on the uniformity and scintillation properties of PbWO$_{4}$. Therefore, the distribution of Y$^{3+}$ is  carefully controlled to ensure uniformity in detector grade crystals~\cite{sic1}.

A list of relatively high density materials and their key physical properties are listed in Table~\ref{tab:alt_mat}. Although having much lower density we have also included silicon, as they could be another choice for a $G$ test mass material in view of the large volumes and very high crystal quality and infrared lasers developed over the past decades for the semiconductor industry. The mass density of PbWO$_{4}$, $\rho $= 8.26~g/cm$^{3}$, is only a factor of 2 smaller than that of tungsten, the densest material commonly used in $G$ measurements. Its transparency opens up the possibility for $G$ experiments to conduct laser interferometric measurements of its dimensions and location, thereby providing a way to cross-check coordinate measuring machine metrology and thereby independently confirm its absolute accuracy. In addition, the existence of such a source mass material might inspire new designs of $G$ apparatus to take advantage of the possibility of in-situ optical metrology to re-optimize the apparatus design tradeoffs between systematic errors and $G$ signal size. In addition to PbWO$_{4}$ there are several additional high density materials that are also optically transparent and could be used as source/test masses.  Neutron and optical interferometric methods can be applied to all of these materials.

\begin{table}[h!]
  \caption{List of non-hygroscopic, optically-transparent, high-density crystals and their physical properties.}
  \begin{center}
    \begin{tabular}{|l|c|c|c|c|c|c|} \hline
      Material & PbWO$_{4}$ & CdWO$_{4}$ & LSO & LYSO & BGO & ~Si~~\\\hline 
      Density [g/cm$^3$] & 8.3 & 7.9 & 7.4 & 7.3& 7.13&2.33\\
      Atomic numbers & 82, 74, 8 & 48, 74, 8 & 71, 32,8 &71, 39, 32,8 & 83, 32, 8 & 14\\
      Refractive index (light) & 2.2 & 2.2-2.3 & 1.82 & 1.82 & 2.15 & 3.45 \\
      & & & & & & (infrared) \\
      Thermal expansion  & 8.3 (para)& 10.2 & 5 & 5 & 7 & 2.6 \\    
      coefficient(s) [10$^{-6}/^{\circ}$C] & 19.7 (perp)& & & & &\\ \hline
    \end{tabular}
  \end{center}  
  \label{tab:alt_mat}
\end{table}  

%\section{}

%\noindent

\section{Experimental Details}
Two 2.3~cm~$\times$~2.3~cm$\times$~12.0~cm PbWO$_{4}$ samples were used in our study and they are shown in Figure~\ref{fig:transp} (right). The PbWO$_{4}$ test masses were initially characterized at Indiana University to confirm that they were geometrically uniform. Visual inspection revealed no evidence for any nonuniform local density anomalies from internal voids or inclusions of the type which might introduce systematic errors in $G$ measurements. The detailed shapes of the four 2.3~cm~$\times$~2.3~cm surfaces for each crystal were measured by the National Institute of Standards and Technology (NIST) Metrology Group on a coordinate measuring machine, with a maximum permissible error (MPE) of $0.3$ $ \mu$m $+ (L/1000)$  $\mu$m where length $L$ is in units of mm. %The angles between the surface normals for the two pairs of crystal faces to be exposed to the laser beam varied between $179.93$ degrees to $179.98$ degrees.

\begin{figure}[hbt!]
  \begin{center}
  \includegraphics[width=8.5cm]{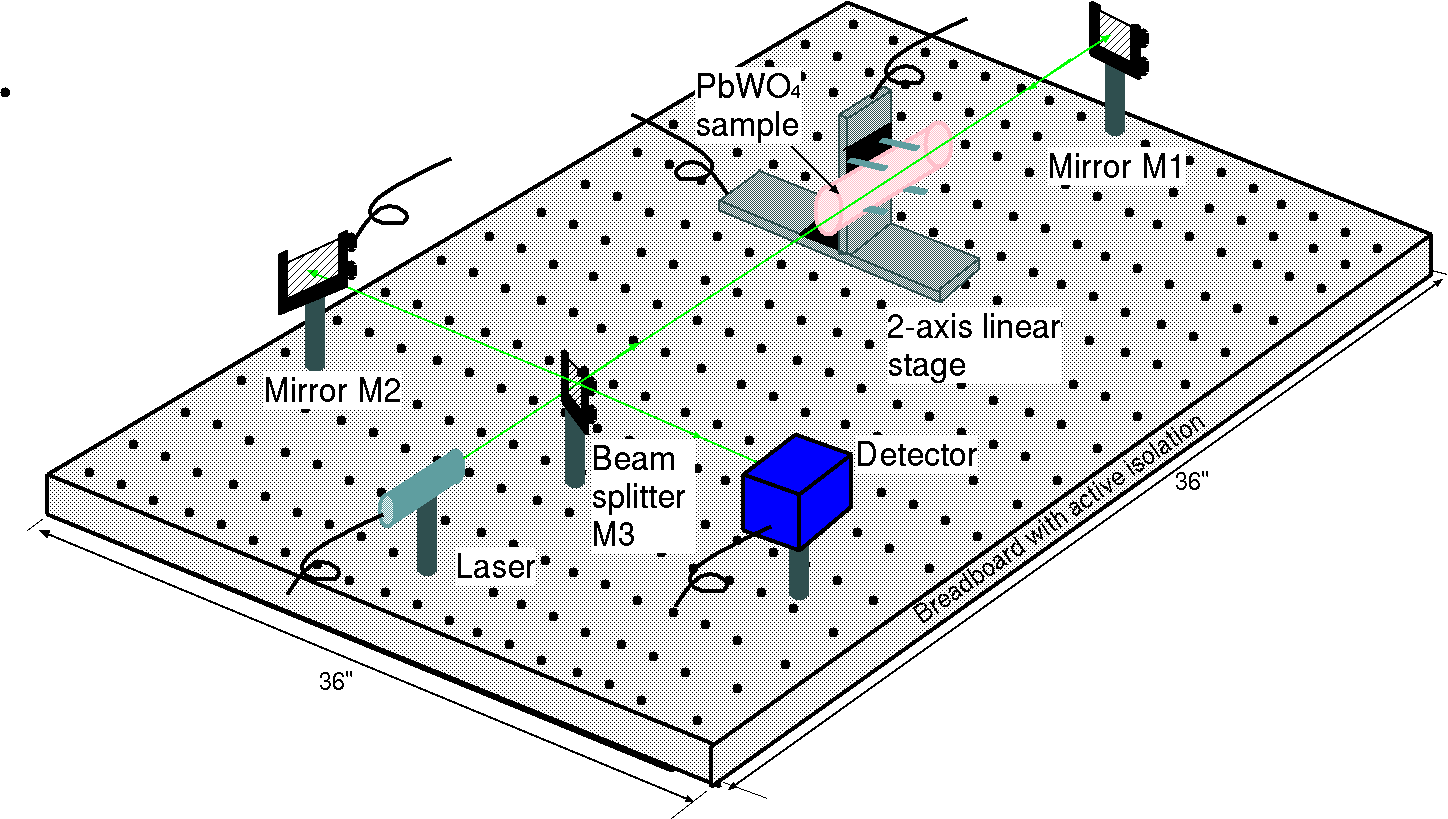}
  \hspace{10ex}
    \includegraphics[width=7.5cm]{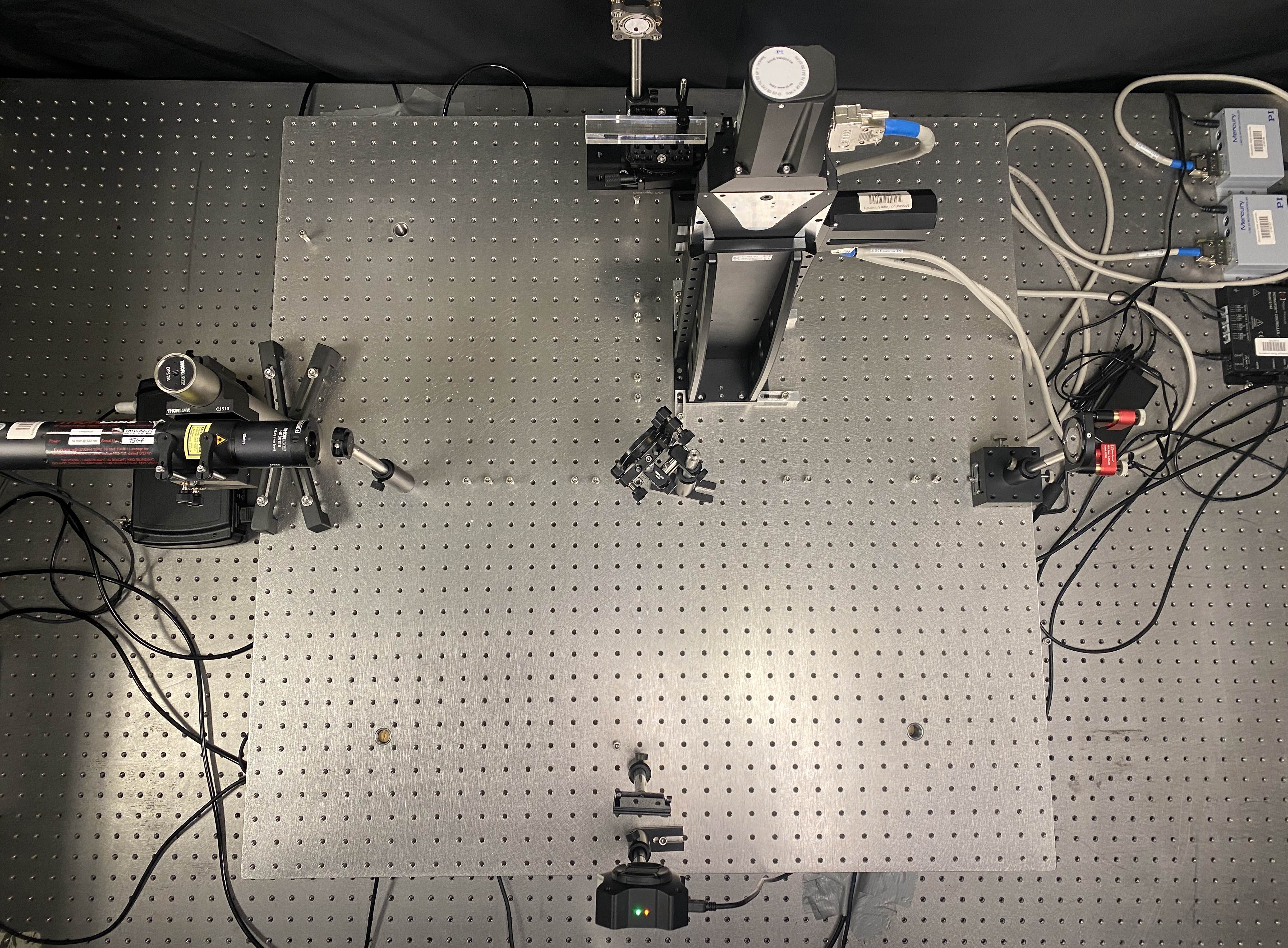}
%   \vspace{50ex} 
  \caption{(left) Schematic diagram of optical interference method of measuring the variation of refractive index. M1 and M2 are reflecting mirrors, and M3 is a beam splitting mirror. (right) Photograph of experimental setup.}
  \label{fig:inter}
\end{center}
\end{figure}

A laser interferometer was built at Mississippi State University (MSU) to measure the internal density inhomogeneities in PbWO$_{4}$. Figure~\ref{fig:inter} (left) shows the schematic of the optical interference method used to measure the variation in the refractive index of the PbWO$_{4}$ blocks. A photograph of the experimental setup is shown in Fig.~\ref{fig:inter} (right).
The light source for the interferometer was a linearly-polarized,  stabilized HeNe Laser of wavelength 632.992~nm, stabilized power of 1.2~mW and beam diameter of 0.65~mm, manufactured by Thorlabs~\cite{thorlabs}. The laser beam was split into two using a Pellicle beam-splitter manufactured by Thorlabs, coated for  50:50 reflection:transmission (R:T) split ratio at 635~nm. Two sets of reflecting laser line mirrors, M1 and M2, obtained from Newport Corporation~\cite{newport} were placed equidistant (44 cm) from the beam-splitter such that the two reflected beams could interfere at the beam-splitter and the interference pattern could be recorded in a CCD camera placed perpendicular to the laser source as shown in Fig.~\ref{fig:inter}.  The mirror M2 was attached to two linear piezoelectric actuators (model number 8816-6) driven by a controller (model number 8742-4-KIT), both from  Newport Corporation. This allowed the mirror to be driven and produce measurable shifts in the fringes to calibrate the interferometer. Two CCD cameras were used; a Thorlabs high-resolution CMOS camera~\cite{cam1}  was used to monitor the interference fringes in 2D while aligning the setup and a Thorlabs smart line camera~\cite{cam2} sensitive over the the range of 350-1100 nm, was used to take images of the fringes in 1D in order to monitoring the shifts in the fringe position.  A neutral density filter was placed in front of the cameras to reduce the intensity of the light hitting the camera. In order to keep the reflected beam from returning to the laser cavity and destabilizing the lasing, a Thorlabs polarization-dependent free-space isolator~\cite{isolator}, was placed between the laser source and the beamsplitter.

\begin{figure}[hbt!]
    \centering
    \subfloat{\centering{{\includegraphics[width=7cm]{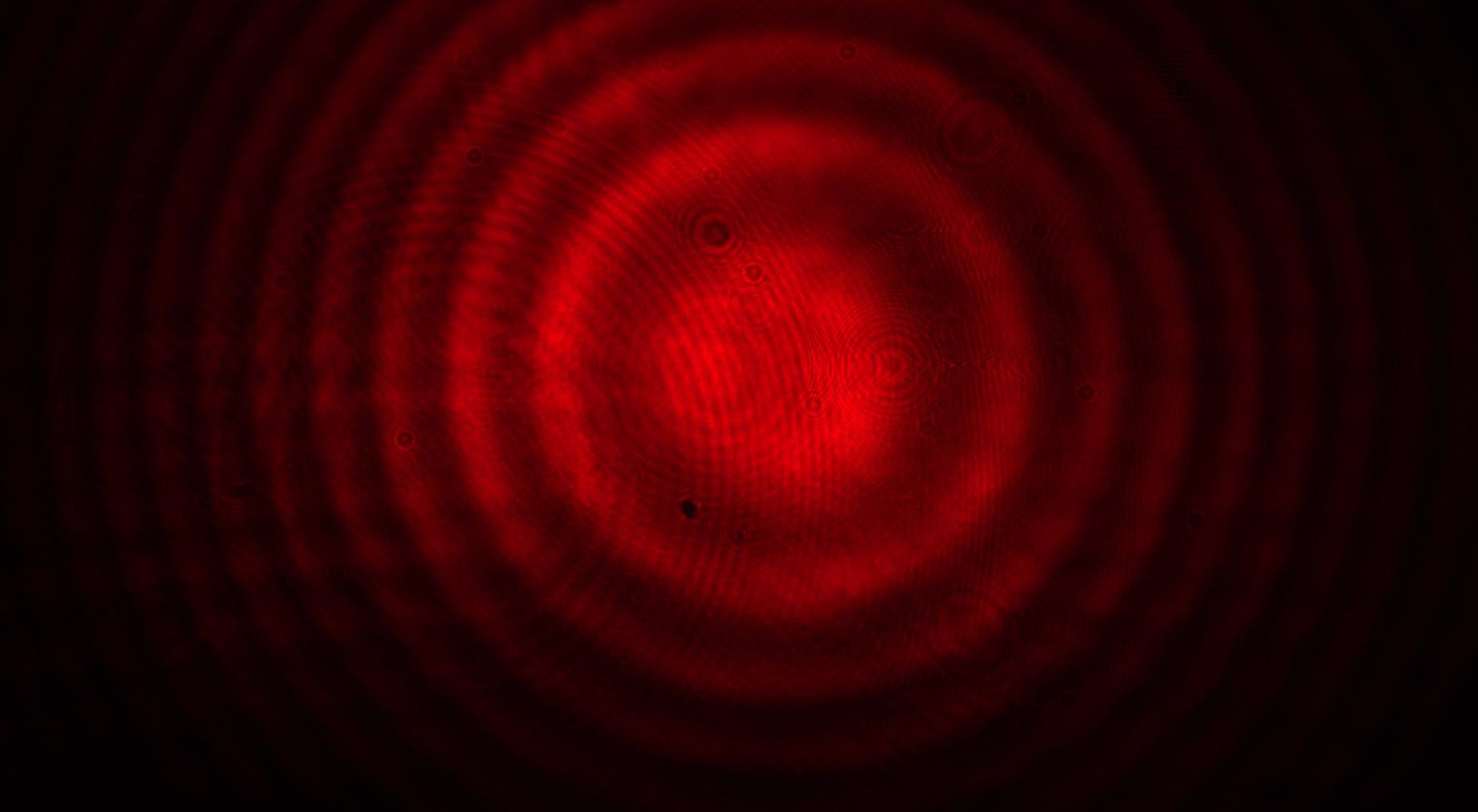}}}}
    %\hspace{5ex}
    \qquad
    \subfloat{\centering{{\includegraphics[width=7.3cm]{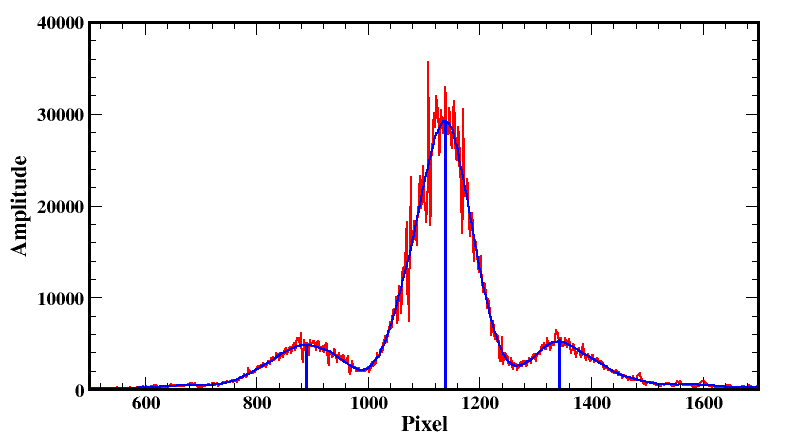}}}}
    \caption{Interference fringes obtained without sample with (left) CC3260C high-resolution USB CMOS camera, and with (right) LC100 smart line camera, respectively. Also, shown are the fits with Gaussian distributions and the position of the centroids of the fringes.}
    \label{fig:figoffringes}
\end{figure}

The PbWO$_{4}$ sample was placed in one of the beam paths (as shown in Fig.~\ref{fig:inter}) on two orthogonally mounted, Physik Instrumente~\cite{PI} 2-axis motorized translation stages (L-509.20DG10 - horizontal axis, L-511.60DG10 - vertical axis). This allowed the entire sample to be scanned by moving it perpendicular to the beam and recording any changes in the interference fringes with the line camera. The entire setup was be assembled on a 36-inch$~\times$~36-inch non-magnetic honeycomb optical breadboard produced by Newport Corporation~\cite{newport1} (SG-33-4-ML) and was supported on a Thorlabs benchtop self-leveling vibration isolation system~\cite{tlab}. The setup was enclosed on light-tight enclosure built using Thorlabs blackout material. The interferometer was controlled with National Instruments LabVIEW~\cite{NI} running on a desktop computer allowing for the automation of the scan of the PbWO$_{4}$ crystals. Fringe tracking was also automated during these scans using fringe skeletonization. 
%The density gradients in two nominally identical samples of PbWO$_4$ crystals were measured. 
Figures ~\ref{fig:figoffringes} (left) and (right) are a 2D and 1D images of fringes, respectively, taken with CMOS Camera and LC100/M line camera.

During the alignment and calibration with the sample out of the beam, a path was defined and the 2-axis motorized translation stage was move in steps of 1~mm while the position of the interference pattern were recorded at each step. The initial and final positions of the peaks were compared and were found to be stable. The motion of the stage and other sources of vibration were found to have no impact on the peak positions. Next, the sample was place in one of the optical beams and the fringes were recorded. The linear stages were moved perpendicular to the beam at 1~mm intervals enabling the scanning of the whole length of the sample while avoiding the edges of the sample. For each of the four faces along the length of the sample, 5 groups of 20 images (100 images per position) are collected in succession, with a delay of 10~s between each group. A 120~s delay is used between each 1~mm step of the linear stage to ensure all vibrations are damped. 

\section{Analysis}
\begin{figure}[hbt!]
  \begin{center}
  \includegraphics[width=12cm]{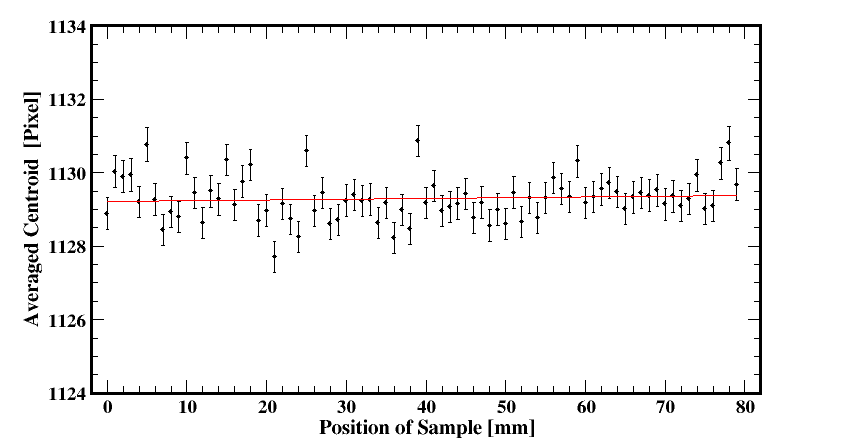}
  \caption{The average centroid versus position for a PbWO$_4$ sample for a typical run. The mean centroid for this run is 1129.04 pixels and the slope is 3.9~$\times$~10$^{-2}$~pixels/cm.}
  \label{fig:avecentroid}
\end{center}
\end{figure}

The density gradients in two nominally identical samples of PbWO$_{4}$ crystals (labeled as MS and IN) were measured using the laser interferometer. The relative shift in the centroid of the central maxima was used to extract the relative change in refractive index which was then converted to density variation and a density gradient. Figure~\ref{fig:figoffringes} (right) shows a typical spectrum from the line camera showing the central maxima and two adjoining fringes. At each position along the length of the crystal, 100 images were recorded with the line camera and the three peaks in the image were fit with a Gaussian distribution to determine the centroids of the peaks. These were then averaged over the 100 images. The average centroid of the central maxima in pixels as a function of the position along the length of the crystal is shown in Figure~\ref{fig:avecentroid}. 

A linear fit of the centroids ($C_i)$ as a function of position was used to determine the relative displacement of the centroid averaged over one complete scan  of the sample along a particular axis ($\Delta C$) and the slope $\frac{\Delta C}{\Delta x}$. The relative displacement and its uncertainty, is given by; 
\begin{equation}
\Delta C = \sqrt{\frac{\sum{_{i=1}^{N} (C - C_{i})^{2}}}{N}},~~~~ \sigma_{\Delta C} = \frac{\Delta C}{\sqrt{N}},
%\end{equation}
%and
%\begin{equation}
~~~~~
\end{equation}
where  $C$ the mean centroid position and $N$ is the total number of position steps. For a typical run $C$ is 1129.04 pixels and $\Delta C$ is 0.83 pixels and the slope is 3.9~$\times$~10$^{-2}$~pixels/cm. The relative displacement of the centroid and its slope is then converted into change in the number of fringes ($\Delta Y$) with an uncertainty ($\sigma_{\Delta Y}$) and gradient $\frac{\Delta Y}{\Delta x}$, given by;
\begin{equation}
\Delta Y = \frac{\Delta C}{W},  ~~~~~\sigma_{\Delta Y} =  
\sqrt{\left( \frac{\sigma_{C}}{C} \right)^{2} + \left( \frac{\sigma_{\Delta C}}{\Delta C} \right)^{2}} \times\Delta Y, ~~~~ \frac{\Delta Y}{\Delta x} = \frac{\Delta C/\Delta x}{W}
\end{equation}
where $W$ is the average width of the central fringe in pixels and $\sigma_{C}$  is the uncertainty of the average centroid position. For a typical run $\Delta Y$ is 9.0$\times$10$^{-3}$ fringes and $\sigma_{\Delta Y}$ is 1.0$\times$10$^{-3}$ fringes and the gradient is 4.2~$\times$~10$^{-4}$~cm$^{-1}$. The variation in the number of the fringes is related to variation in refractive index and its gradient as ~\cite{Liu2008},

\begin{equation}
\frac{\Delta n}{n} = \frac{\lambda \Delta Y }{2 n d},~~~~ \frac{1}{n}\frac{\Delta n}{\Delta x} = \frac{\lambda}{2 n d}\frac{\Delta Y }{\Delta x},
\end{equation}
where $n$ and $\Delta n$ are the refractive index of the sample and its variation, $\lambda$ is the wavelength of the laser, and $d$ is the thickness of sample. For our experiment, $n$ = 2.2,  $\lambda$ = 632.991 nm, and $d$ = 2.3 cm along both x- and y-axes. The average values for $\frac{\Delta n}{n}$ for the MS and IN samples over 38 and 55 runs respectively, are; 

\begin{equation}
\begin{aligned}
\bigg \langle \frac{\Delta n}{n} \bigg \rangle_{MS} = (3.8 \pm 3.6)\times 10^{-8}, ~~~~
\bigg \langle \frac{\Delta n}{n} \bigg \rangle_{IN} = (3.9 \pm 3.6)\times 10^{-8},
\end{aligned}
\end{equation}
and the average values of the gradient for samples are ;
\begin{equation}
\begin{aligned}
\bigg \langle \frac{1}{n}\frac{\Delta n}{\Delta x} \bigg \rangle_{MS} = (5.6 \pm 5.6)\times 10^{-9} {\mbox{cm}}^{-1}, ~~~~
\bigg \langle \frac{1}{n}\frac{\Delta n}{\Delta x} \bigg \rangle_{IN} = (5.6 \pm 3.9)\times 10^{-9} {\mbox{cm}}^{-1}.
\end{aligned}
\end{equation}

The relation between changes in refractive index and variation in density for transparent materials is governed by the Gladstone-Dale empirical linear equation, established in previous characterization studies \cite{MajS1884,MarlerB1988,ScholzeH1977,HugginsML1943,MandarinoJA1976}, 
\begin{equation}\label{eq:gladstone_dale}
%\bar{n} = r\bar{\rho} + 1, ~~~r = \frac{1}{100}{\sum {r_{i} p_{i}}}, ~~~~\frac{1}{\bar{\rho}} = \frac{1}{100}{\sum{S_{i} p_{i}}},
\bar{n} = r\bar{\rho} + 1,
\end{equation}
where $r$ is the specific refractive index energy, $\bar{n}$ and $\bar{\rho}$ are the average of the refractive index and density of the sample respectively. For a multi-component material, specific refractive index energy $r$ and average density $\bar{\rho}$ depends on the chemical composition of the individual components according to the additive law \cite{ScholzeH1977,HugginsML1943,MandarinoJA1976},
%\begin{equation}\label{eq:ave_r}
 %\end{equation}
%\begin{equation}\label{eq:ave_rho}
%\end{equation}
\begin{equation}\label{eq:additive}
r = \frac{1}{100}{\sum {r_{i} p_{i}}}, ~~~~\frac{1}{\bar{\rho}} = \frac{1}{100}{\sum{S_{i} p_{i}}},
\end{equation}
with $r_{i}$, $S_{i}$ and $p_{i}$ as the specific refractive energy, density factor and weight percentage of the individual components respectively. The parameters for the components of PbWO$_{4}$ are listed in Table~\ref{Tab:pbwo4comp}.

\begin{table}[htb]
%\centering
\caption{Parameters for components of PbWO$_{4}$}
\begin{center}
\begin{tabular}{|l|l|l|l|}\hline
 Components & ~~~$S_{i}$~~~ & ~~~$r_{i}$~~~ & ~~~$p_{i}$(\%)~~~\\\hline
 PbO &  0.0926 & 0.1272 &  ~~49.05\\
 WO$_{3}$ & 0.2826* & 0.1420 & ~~50.95\\\hline
\end{tabular}
\end{center}
%\caption{Parameters for components of PbWO$_{4}$}
\label{Tab:pbwo4comp}
\end{table}

The density factor for WO$_{3}$ was not found in the literature, therefore, we calculate it using the Gladstone-Dale relationship \cite{Bloss1982}
\begin{equation}\label{eq:dencomp}
K = \frac{n-1}{\rho} = \frac{1}{100}{\sum {r_{i} p_{i}}},
\end{equation}
where $K$ is the specific refractive index energy of material with refractive index $n$ and density $\rho$, and $r_{i}$ and $p_{i}$ are the refractive energy constants and weight percentages for the constituents in the compound, respectively. From Eq.~\ref{eq:additive} the relative variation of density and specific refractive index energy to first order is;
\begin{equation}\label{eq:rapprx}
\frac{\Delta r}{r} \approx \frac{1}{M} \sum_{n=1}^{n} \left(\frac{r_{i}}{r} - 1 \right) \Delta m_{i},
%\end{equation}
%\begin{equation}\label{eq:denapprx}
~~~~\frac{\Delta \rho}{\rho} \approx \frac{1}{M} \sum_{n=1}^{n} (1-S_{i} \bar{\rho}) \Delta m_{i},
\end{equation}
where $M$ is the mass of the sample, $\Delta m$ the mass fluctuations of an arbitrary component in the sample. From Eq.~\ref{eq:gladstone_dale}~and~\ref{eq:rapprx} 
%and \ref{eq:denapprx}, 
we get;
\begin{equation}\label{eq:finalans}
%\frac{\Delta \rho}{\rho} \approx 
\Bigg \langle \left(\frac{\Delta \rho}{\bar{\rho}} \right)^{2} \bigg \rangle = \left(1 + \frac{1}{r \bar{\rho}}\right)^{2} \frac{\sum_{i=1}^{n} (1-S_{i} \bar{\rho})^{2}}{\sum_{i=1}^{n} (\frac{r_{i}}{\bar{r}} - S_{i} \bar{\rho})^{2}} \Big \langle \Big( \frac{\Delta n}{n}\Big )^{2} \Big \rangle
\end{equation}
where $\bar{r}$ is the is a simple average of $r$ of individual components. Substituting the values of $S_{i}$, $r_{i}$ and $p_{i}$ in table \ref{Tab:pbwo4comp} into Eq.~\ref{eq:finalans}, we obtain $\frac{\Delta \rho}{\bar{\rho}}~=~3.8212 \frac{\Delta n}{\bar{n}}$. Further, substituting the measured value of $\frac{\Delta n}{\bar{n}}$ and its gradient we obtain the average density variation of;
 \begin{equation}
 %\left(\frac{\Delta \rho}{\bar{\rho}}\right)_{MS}~=~(1.4 \pm 1.4) \times 10^{-7}~~~~\left(\frac{\Delta \rho}{\bar{\rho}}\right)_{IN}~=~(1.5 \pm 1.4) \times 10^{-7}, 
 \bigg \langle \frac{\Delta \rho}{\bar{\rho}} \bigg \rangle_{MS}~=~(1.4 \pm 1.4)\times 10^{-7}, ~~~~\bigg \langle \frac{\Delta \rho}{\bar{\rho}} \bigg \rangle_{IN}~=~(1.5 \pm 1.4)\times 10^{-7},  
 \end{equation}
 and a density gradient of;
\begin{equation}
\bigg \langle \frac{1}{\bar{\rho}}\frac{d \rho}{dx} \bigg \rangle_{MS}~=~(2.1 \pm 2.2) \times 10^{-8}~~ {\mbox{cm}}^{-1}, ~~~~\bigg \langle \frac{1}{\bar{\rho}}\frac{d \rho}{dx} \bigg \rangle_{IN}~=~(2.1 \pm 1.5) \times 10^{-8}~~ {\mbox{cm}}^{-1}
\end{equation}
for the two samples. These results are consistent but $\sim$ 4 times more sensitive than the neutron phase contrast imaging measurements of these same samples~\cite{neutron}. 

\section{Conclusions and Future Work}
Two PbWO$_{4}$ crystals were tested using a newly constructed laser interferometer. The density inhomogeneity and gradient was quantified in terms of variation in the interference fringes as a laser beam scanned the crystals. A measured inhomogeneity upper-limit of $\frac{\Delta \rho}{\bar{\rho}}~<~1.4 \times 10^{-7}$ and a density gradient upper-limit of $\frac{1}{\bar{\rho}}\frac{d \rho}{dx}~<~2.1 \times 10^{-8}~~ {\mbox{cm}}^{-1}$ was obtained from the two PbWO$_{4}$ crystals. The density gradient is about a factor of 4 better than an earlier measurements using a neutron Talbot-Lau interferometer but still consistent with the order-of-magnitude estimates based on the level of impurities in commercially-available PbWO$_4$ crystals. The size of the density gradient is also consistent with the NIST coordinate measuring machine measurements of the small deviations of the shape of the crystals away from our assumed geometry. These measured density inhomegeneity and gradient
is more than three orders of magnitude smaller than what is needed in typical macroscopic mechanical apparatus of the type often used to measure $G$. We conclude that PbWO$_{4}$ is a strong candidate for use as test/source masses in future $G$ measurements. 

The same measurement methods used in this work would also work for many other optically (infrared) transparent candidate $G$ test mass materials.  If further independent confirmation of the results is desired, another probe which could penetrate the required thickness of material is needed. This can be done using traditional gamma ray transmission radiography. A gamma ray radiography facility at Los Alamos can routinely penetrate several centimeters of dense material and resolve internal voids at the millimeter scale~\cite{Espy}.  All three of these methods (neutron and optical imaging and gamma transmission) are nondestructive and can therefore in principle be applied to the same $G$ test mass used in the actual experimental measurements. Future measurements of internal density gradients of test masses used for measurements of $G$ can now be conducted for a wide range of possible test masses in a nondestructive manner. We are therefore encouraged to think that we have established a method which can address one of the common potential systematic error sources in measurements of $G$.

\section{Acknowledgements}
We would like to thank V. Lee, D. Newell and S. Schlamminger of the National Institute of Standards and Technology in Gaithersburg, MD for their help in arranging the coordinate measuring machine work. K. T. A. Assumin-Gyimah, D. Dutta, and W. M. Snow acknowledge support from US National Science Foundation grant PHY-1707988. W. M. Snow acknowledges support from US National Science Foundation grants PHY-1614545 and PHY-1708120 and the Indiana University Center for Spacetime Symmetries. M. G.~Holt was supported by the Department of Physics and Astronomy at Mississippi State University. 

%\section{Appendix}

\end{document}